\pdfoutput=1
\documentclass[11pt]{article}

\usepackage[final]{acl}

\usepackage{times}
\usepackage{latexsym}

\usepackage[T1]{fontenc}
\usepackage[utf8]{inputenc}

\usepackage{microtype}

\usepackage{inconsolata}

\usepackage{graphicx}

\usepackage{bbm}
\usepackage{subcaption}
\usepackage{multirow}
\usepackage{enumerate}
\usepackage{amsmath}
\usepackage{microtype}
\usepackage{amsmath}
\usepackage{amsthm,amssymb}
\usepackage{graphicx}
\usepackage{booktabs} % for professional tables
\usepackage{multirow}

\usepackage{array}
\usepackage{arydshln}
\usepackage{color}
\usepackage{xcolor}

\title{Modeling News Interactions and Influence for Financial Market Prediction}

\author{
 \textbf{Mengyu Wang} \quad
 \textbf{Shay B. Cohen} \quad
 \textbf{Tiejun Ma}
\\
 School of Informatics, The University of Edinburgh,
\medskip
\\
 \texttt{\{mengyu.wang,scohen,tiejun.ma\}@ed.ac.uk}
}

\newenvironment{enumeratesquish}[2]{\begin{list}{\labelenumi}{\setlength{\itemsep}{#1}\setlength{\labelwidth}{#2}\setlength{\leftmargin}{\labelwidth}\addtolength{\leftmargin}{\labelsep}}}{\end{list}}

\begin{document}
\maketitle
\begin{abstract}
The diffusion of financial news into market prices is a complex process, making it challenging to evaluate the connections between news events and market movements. This paper introduces FININ (Financial Interconnected News Influence Network), a novel market prediction model that captures not only the links between news and prices but also the interactions among news items themselves. FININ effectively integrates multi-modal information from both market data and news articles. We conduct extensive experiments on two datasets, encompassing the S\&P 500 and NASDAQ 100 indices over a 15-year period and over 2.7 million news articles. The results demonstrate FININ's effectiveness, outperforming advanced market prediction models with an improvement of 0.429 and 0.341 in the daily Sharpe ratio for the two markets respectively. Moreover, our results reveal insights into the financial news, including the delayed market pricing of news, the long memory effect of news, and the limitations of financial sentiment analysis in fully extracting predictive power from news data.
\end{abstract}

\section{Introduction}
\label{introduction}

Media coverage and news events have been shown to influence market returns~\cite{tetlock2007giving, sattler2013markets}. However, the process through which news diffuses into market prices is complex. The influence of news items on market outcomes varies, affected by various market conditions~\cite{cheung2019exchange, hirshleifer2022macro}. Since existing studies often adopt a simplified approach by treating available news data holistically and investigating its overall effect on the market~\cite{wang2018combining,lopez2023can}, the nuanced information contained within individual news items is overlooked. Consequently, parsing the market information diffusion process in detail by considering the impact of each news item emerges as a potential approach to enhance market prediction.
\begin{figure}[t]
% \vspace{-0.6cm}
\centering
\begin{center}
   \includegraphics[width=1.0\linewidth]{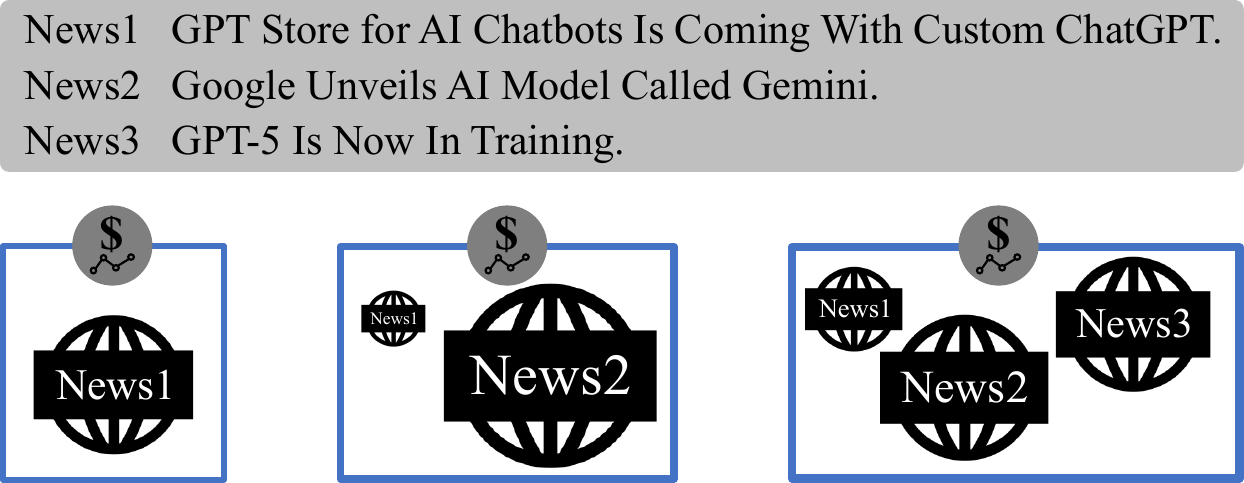}
    \vspace{-0.6cm}
   \captionof{figure}{News importance fluctuates when considered in the context of other news. Breaking news (News 2) may diminish the market impact of another news item (News 1). However, if related breaking news (News 3) is considered, the influence of News 1 could resurge.}
\label{fig:news_impact_example}
\end{center}
\vspace{-0.6cm}
\end{figure}

However, evaluating the influence of individual news items is challenging. News does not exist in isolation but is connected with other information. Studies indicate that information diffusion within the market is a gradual process~\cite{albert1996market, kerl2007market}. Financial news exhibits a ``long memory effect'', a persistent impact over time~\cite{ho2013does}. During this dissemination process, the influence of an individual news item can be further modulated by interactions with other news items, as illustrated by the practical example in Figure~\ref{fig:news_impact_example}. Given the sheer volume of news, evaluating the market impact of news items becomes even more challenging.

This paper aims to address this gap. The domain of financial news presents an excellent testbed for not only improving market prediction but also enhancing our understanding of how to effectively leverage \textbf{multi-modal data}. Modeling news interactions requires a more comprehensive use and exploration of multi-modal data in the form of both numerical and textual data from market and news sources. It also requires considering established market theories (see \S\ref{sec:related_work}), such as the \textbf{efficient market hypothesis} and \textbf{information diffusion theory}.

We develop \textbf{FININ}, a \textbf{F}inancial \textbf{I}nterconnected \textbf{N}ews \textbf{I}nfluence \textbf{N}etwork to build interactions between news reports and analyze their diverse impact on markets. The model comprises two key components: a data fusion encoder and a market-aware influence quantifier. The data fusion encoder handles multi-modal data by encoding various inputs differently while maintaining consistency for data within the same domain. It encodes information related to each news report as independent features, providing a foundation for modeling complex information interactions.
The market-aware influence quantifier draws upon financial theories to analyze relationships both between news items themselves and between news and prices. It evaluates the market influence of each news item by considering the overall market context.

Experiments on two major markets, S\&P 500 and NASDAQ 100, demonstrate the superiority of FININ over state-of-the-art methods. Using a GPT-based method~\cite{lopez2023can} as a baseline, we show that large language models (LLMs) cannot be directly used for managing large volumes of news for market prediction. While LLMs perform well on financial sentiment analysis, FININ demonstrates a better ability to build relationships between news and market movements. 

Additionally, our results provide three key insights into the market information diffusion process. Firstly, FININ highlights a delay in the market pricing of news, indicating market inefficiency in incorporating news information. Secondly, we corroborate the long memory effect of news~\cite{lillo2015news, ho2013does}. Thirdly,
we identify the limitations of financial news sentiment analysis and describe how FININ overcomes these limitations. These findings enhance our understanding of the market's response to news.
To summarize, our main contributions are:
\begin{enumeratesquish}{0em}{0.5em}
    \item[1] We propose FININ, a multi-modal model that quantifies the market impact of individual news items by considering the interactions among news and the links between news and prices.
    \item[2] The FININ model demonstrates its effectiveness in leveraging news information for market prediction, outperforming state-of-the-art methods with improvements of at least 0.429 and 0.341 in the daily risk adjusted returns (Sharpe ratio) for the S\&P 500 and NASDAQ 100 respectively.
    \item[3] FININ's findings have revealed three key insights: the delayed market pricing of news, the long memory effect of news, and the limitations of financial news sentiment analysis in extracting predictive power from news data.
\end{enumeratesquish}

\section{Related Work}
\label{sec:related_work}

\textbf{News-based Market Prediction.} 
Financial news has been proven to significantly affect market movements, impacting investor sentiment and decision-making~\cite{tetlock2007giving, barber2008all, calomiris2019news}. Advancements in data mining and natural language processing (NLP) have enabled the analysis of news data, including tweets and online news, for market prediction~\cite{dewally2003internet, li2010information, xu2018stock}. The extraction of events and sentiments from news text has progressed substantially~\cite{xu2021rest, hu2019transformation}. Furthermore, progress in deep learning has led to the development of effective neural network structures, such as multi-modal LSTM models~\cite{li2020multimodal, dong2020belt}, ensemble models~\cite{li2022novel}, and hierarchical attention models~\cite{luo2023causality}, which leverage news sentiments for stock prediction.
However, existing research often focuses on the market influence of overall news, neglecting the nuanced information and varying impacts of individual news items. While some studies recognize the diverse relationships between news and prices~\cite{huynh2017stock} and attempt to quantify the varying market impact of news~\cite{wang2024mana}, fewer have explored the interactions between multiple news items. Additionally, more complex aspects related to the news impact on market, such as the information diffusion process~\cite{bekiros2017information}, are not fully considered during the prediction process.

\noindent\textbf{Market Information Diffusion.} 
The Efficient Markets Hypothesis (EMH) posits that stock prices instantaneously reflect all available information, leaving no room for excess returns~\cite{fama1970efficient}. However, real-world complexities challenge this ideal situation. In contrast, the information diffusion hypothesis (IDH) suggests that news-induced information is gradually incorporated into prices, leading to delayed market reactions~\cite{kerl2007market, zhang2016market}. Research highlights the enduring impact of news on market prices over an extended period~\cite{ray2000long, christensen2007effect}, requiring the analysis of the relationship between news and price changes. Moreover, news items are not isolated in their effect on the market but are interconnected, with their market impact influenced by one another~\cite{yu2019online, agarwal2019stock}. Therefore, effectively capturing the complexities of the information diffusion process remains a challenge, requiring further research to develop models that capture comprehensive and detailed market impact of news items.

\section{Preliminaries}
\subsection{Notation and Problem Formulation}
\label{sec:formulation}
In our approach, we define $d$ as a trading day. Our data consists of daily market data $\boldsymbol{M}_d$ and news report data $\boldsymbol{R}_d$. Both data sources contain textual information (denoted by $*^{(t)}$) and numerical information (denoted by $*^{(n)}$). The market data, $\boldsymbol{M}_d = (\boldsymbol{m}^{(t)}_{d},  \boldsymbol{m}^{(n)}_{d})$, consists of textual information $\boldsymbol{m}^{(t)}_{d}$, including market descriptions and a list of most related companies, and numerical information $\boldsymbol{m}^{(n)}_{d}$, comprising daily market prices. The daily news data $\boldsymbol{R}_d = (\boldsymbol{R}_{d, 1}, \boldsymbol{R}_{d, 2}, ..., \boldsymbol{R}_{d, N_d})$ contains multiple news reports, where each $\boldsymbol{R}_{d, i}$ represents information associated with the $i$-th news item on day $d$, and $N_d$ represents the news volume on day $d$. Each news report $\boldsymbol{R}_{d, i} = (\boldsymbol{r}^{(t)}_{d, i},  \boldsymbol{r}^{(n)}_{d, i})$ comprises textual information $\boldsymbol{r}^{(t)}_{d, i}$, which includes the news headline, and numerical information $\boldsymbol{r}^{(n)}_{d, i}$, representing the news market sentiment scores.

This paper focuses on daily market prediction. The target variable $y_d$ represents the daily market trend, defined as a binary value where 1 signifies a price increase and 0 indicates no change or a decrease between day $d$ and day $d+1$. Our objective is to effectively use the information from the prior $t$ days up to day $d$, including both market data $(\boldsymbol{M}_{d-t+1}, ..., \boldsymbol{M}_{d-1}, \boldsymbol{M}_d)$ and news data $(\boldsymbol{R}_{d-t+1}, ..., \boldsymbol{R}_{d-1}, \boldsymbol{R}_d$), to predict $y_d$.

\subsection{Data}
\label{sec:data}
We use two major stock market datasets: the Standard and Poor's 500 (S\&P 500) Index and the NASDAQ 100 Index. Additionally, we use the Thomson Reuters News Analytics (TRNA) dataset, comprising over 2.7 million news items. The volume of our news data is significantly larger, at least 10 times greater, than those used in previous studies~\cite{mohan2019stock, lopez2023can, luo2023causality}. Furthermore, unlike their online-collected news data, our dataset is sourced from industry practice, ensuring higher quality and more comprehensive market information. Detailed information about these datasets is presented in \S\ref{sec:app_data}

We collected daily S\&P 500 and NASDAQ 100 prices from 2003 to 2018 as market numerical data $\boldsymbol{m}^{(n)}_{d}$. Also, we collected the descriptions of these two markets and their constituent companies' names as textual data $\boldsymbol{m}^{(t)}_{d}$. For TRNA dataset, we consider headlines as news textual data $\boldsymbol{r}^{(t)}_{d, i}$ and the sentiment scores as news numerical data $\boldsymbol{r}^{(n)}_{d, i}$.

\section{Method}
\label{sec:method}

\begin{figure*}[t]
\centering
\begin{center}
   \includegraphics[width=0.95\linewidth]{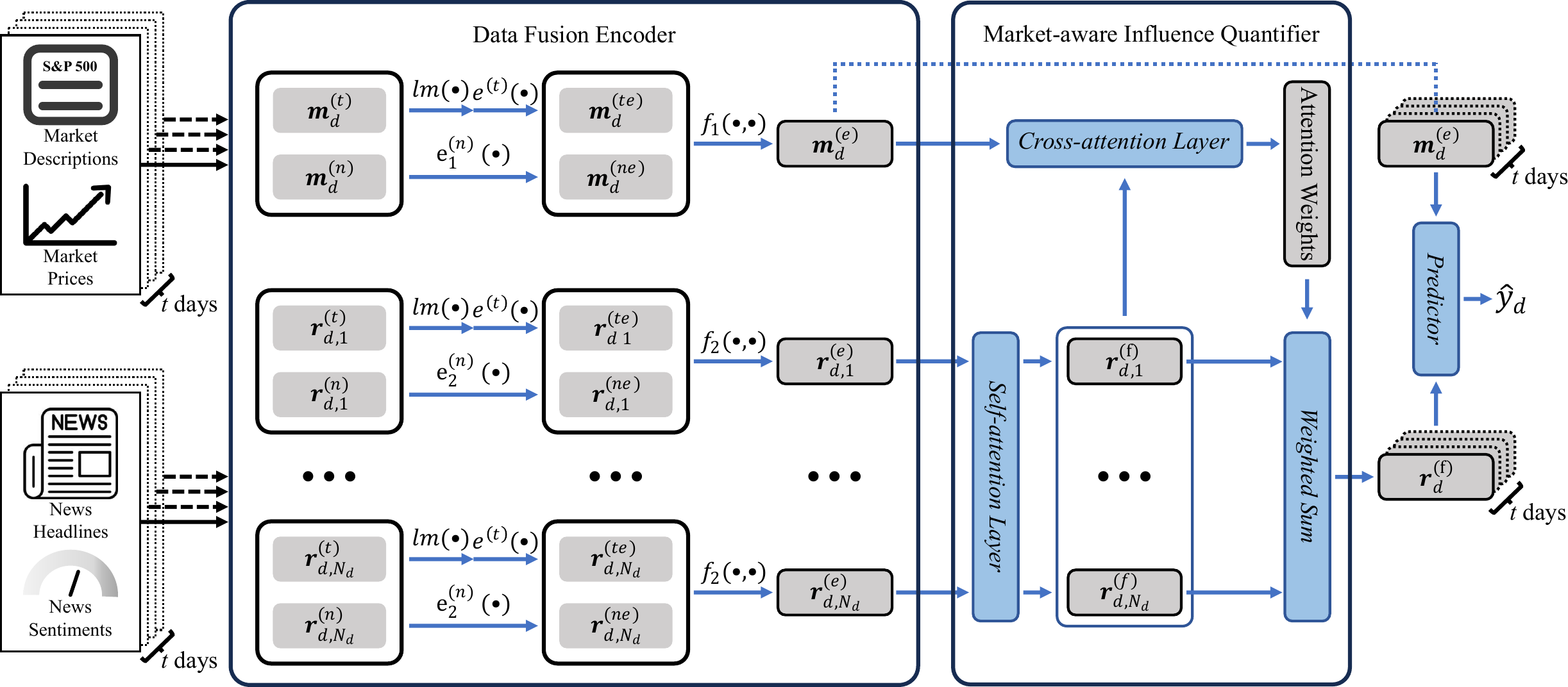}
   \vspace{-0.1cm}
   \captionof{figure}{An overview of the FININ model. It comprises two components: a Data Fusion Encoder and a Market-aware Influence Quantifier. The annotations are defined and explained in the corresponding parts of Section~\ref{sec:method}}
\label{fig:combined_model}
\end{center}
\vspace{-0.7cm}
\end{figure*}

We have designed the FININ model to evaluate the impact of news on price changes. The FININ model comprises two primary components: a data fusion encoder and a market-aware influence quantifier. By leveraging both textual and numerical data, FININ can capture the mutual influence between these disparate pieces of information. Figure~\ref{fig:combined_model} illustrates an overview of the FININ model.

\subsection{Data Fusion Encoder}
Existing news-based market prediction methods primarily focus on the relationship between overall news and price changes, leveraging either market sentiment analysis~\cite{wang2018combining, jing2021hybrid, lopez2023can} or news text processing~\cite{xu2018stock, luo2023causality}. However, these approaches have two main limitations. First, sentiment analysis alone may fail to capture nuanced semantic information in the news text, while text processing may lack insights into the market influence conveyed by sentiment scores. Second, these methods treat all news for one prediction as a single entity, potentially overlooking the varying relationships between individual news items and market movements.

In contrast, our proposed FININ model treats each news report as a distinct unit and uses both types of data, capturing more comprehensive market-related information conveyed by individual news reports. The data fusion encoder tackles the multi-modal inputs by encoding various data types from market conditions and news items into meaningful features. It first encodes inputs into features individually and then integrates features from the same input unit (market snapshot or news item) into fusion features.

The core idea of our encoder is to design encoding functions tailored to the nature of each data source, ensuring consistent and differentiated processing. For textual data from market descriptions $\boldsymbol{m}^{(t)}_{d}$ and news headlines $\boldsymbol{r}^{(t)}_{d, i}$, we use the same encoding function to create unified semantic representations across different text sources. We leverage a pre-trained language model (LM), denoted by $\textit{lm}(\cdot)$, coupled with an encoding layer $e^{(t)}(\cdot)$, to extract market and news text features $\boldsymbol{m}^{(te)}_{d}$ and $\boldsymbol{r}^{(te)}_{d, i}$, respectively:
% \begin{equation}
\begin{align}
\label{eq:text_encoding}
\boldsymbol{m}^{(te)}_{d} &= e^{(t)} (\textit{lm}(\boldsymbol{m}^{(t)}_{d} )), \\
\boldsymbol{r}^{(te)}_{d, i} &= e^{(t)} (\textit{lm}(\boldsymbol{r}^{(t)}_{d, i})).
\end{align}
% \end{equation}
The pre-trained LM is frozen, while the added layer $e^t(\cdot)$ adapts the representations for our prediction task. We explore various LMs for financial text, which will be discussed in detail in \S~\ref{sec:experiment}.

For numerical data, due to their different information content and input sizes, market data $\boldsymbol{m}^{(n)}_{d}$ and news sentiment scores $\boldsymbol{r}^{(n)}_{d, i}$ are encoded using separate functions $e^{(n)}_1(\cdot)$ and $e^{(n)}_2(\cdot)$, respectively. This ensures tailored processing while maintaining consistency in encoding numerical data from different news items. The resulting features are denoted by $\boldsymbol{m}^{(ne)}_{d}$ and $\boldsymbol{r}^{(ne)}_{d, i}$ for market and news data, respectively, as shown in the equation:
\begin{equation}
\label{eq:non-text_encoding}
\boldsymbol{m}^{(ne)}_{d} = e^{(n)}_1 (\boldsymbol{m}^{(n)}_{d}), \boldsymbol{r}^{(ne)}_{d, i} = e^{(n)}_2(\boldsymbol{r}^{(n)}_{d, i}).
\end{equation}

To integrate textual and numerical information into unified encodings, we use separate data fusion functions $f_1(\cdot, \cdot)$ and $f_2(\cdot, \cdot)$ for market and news data, respectively. Similar to the processing of numerical data, these functions ensure consistent encoding within news items while distinguishing between the two sources. The functions $f_1(\cdot, \cdot)$ and $f_2(\cdot, \cdot)$ are implemented as a concatenation of the textual and numerical input features, followed by a multi-layer perceptron (MLP). They take the textual and numerical features as inputs and produce the integrated encodings $\boldsymbol{m}^{(e)}_{d}$ and $\boldsymbol{r}^{(e)}_{d, i}$ for market and news items, respectively:
\begin{align}
\label{eq:data_fusion}
\boldsymbol{m}^{(e)}_{d} &= f_1 (\boldsymbol{m}^{(te)}_{d}, \boldsymbol{m}^{(ne)}_{d}), \\
\boldsymbol{r}^{(e)}_{d, i} &= f_2 (\boldsymbol{r}^{(te)}_{d, i}, \boldsymbol{r}^{(ne)}_{d, i}).
\end{align}

By using the multi-modal data as inputs, we obtain features containing more comprehensive information. Moreover, information relevant to each news report is encoded into independent features, enabling an analysis of individual news items.

\subsection{Market-aware Influence Quantifier}
Unlike existing works that treat news data holistically to obtain daily features, FININ uses independent features for each news item to capture market insights. This section describes the Market-aware Influence Quantifier, which leverages financial theories to better reflect real-world market dynamics.

Firstly, the market information diffusion theory~\cite{kerl2007market, zhang2016market} suggests that information pieces influence each other. To capture these interactions, we use an attention layer~\cite{vaswani2017attention} to refine news encodings $\boldsymbol{r}^{(e)}_{d, i}$ into context-aware features $\boldsymbol{r}^{(f)}_{d, i}$. These context features still focuses on the $i$-th news of day $d$ but incorporate information from the entire daily news corpus, making each news report ``aware'' of the others. 

Secondly, the efficient market hypothesis posits that all relevant information is reflected in market prices. To account for this connection, we introduce another attention layer. In contrast to the first layer that builds relationships between all input states, this layer focuses solely on the connections between the market state and news items. In this layer, the market information unit $\boldsymbol{m}^{(e)}_{d}$ acts as the query, and all news items serve as key states. We calculate query-key attention scores $atts_{d, i}$ to quantify the relative importance of each news report in relation to the market:
\begin{equation}
\label{eq:att1}
atts_{d, i} = \frac{1}{\sqrt{d_k}}\cdot q(\boldsymbol{m}^{(e)}_d) \cdot k(\boldsymbol{r}^{(f)}_{d, i}).
\end{equation}
where $d_k$ denotes the key vector dimension.

These scores are then normalized using a softmax function to obtain attention weights. Notably, we exclude the self-attention score for the market information unit $\boldsymbol{m}^{(e)}_{d}$, ensuring that the resulting weights correspond only to news states and sum to one. Finally, we use these weights to create a weighted sum of news features $\boldsymbol{r}^{(f)}_{d, i}$, aggregating them into overall daily news features $\boldsymbol{r}^{(f)}_{d}$.

By using these two attention layers, we establish relationships between news items themselves and between news and market information. This allows the information states to capture a more comprehensive view of the market, leading to more informative news representations.

\subsection{Market Prediction}
After the data encoding and market influence quantifying stages, the final step of FININ is market prediction. We collect the encoded market and news representations for the past $t$ days, denoted as $(\boldsymbol{m}^{(e)}_{d-t+1}, ..., \boldsymbol{m}^{(e)}_d)$ and $(\boldsymbol{r}^{(f)}_{d-t+1}, ..., \boldsymbol{r}^{(f)}_d)$), respectively. These representations are fed into a prediction structure, denoted as $pred(\cdot, \cdot)$, which aims at forecasting market trends:
\begin{equation}
\begin{split}
\hat{y}_d = \textit{pred}(&(\boldsymbol{m}^{(e)}_{d-t+1}, ..., \boldsymbol{m}^{(e)}_d), \\
&(\boldsymbol{r}^{(f)}_{d-t+1}, ..., \boldsymbol{r}^{(f)}_d)).
\end{split}
\end{equation}

The predictor architecture in FININ is flexible, allowing integration with various widely used stock prediction structures, such as Multi-Layer Perceptron (MLP), Convolutional Neural Network (CNN) and Long Short-Term Memory (LSTM) networks~\cite{jing2021hybrid,jiang2021applications}. Our experiments suggest that the informative features generated by FININ enhance performance regardless of the specific structure. Given the frequent use of MLPs, as subsequent layers due to their simplicity and efficiency~\cite{orimoloye2020comparing, hu2020hrn}, we adopt an MLP for constructing the final FININ model to maintain simplicity and effectiveness.

\section{Experiment Settings}
\label{sec:experiment}

\subsection{Language Models}
\label{sec:benchmark_lm}
Financial language differs from general text due to its area-specific vocabulary and domain knowledge~\cite{araci2019finbert}. While numerous pre-trained LMs exist, their performance on general tasks may not accurately reflect their effectiveness for financial tasks. To investigate how different LMs affect the modeling of market information interactions, we experiment with four different types of LMs within the FININ model for textual data encoding. These LMs include \noindent\textbf{RoBERTa}, \noindent\textbf{FinBERT},  \noindent\textbf{BGE}, \noindent\textbf{Llama2}, representing pre-trained encoder models, domain-specific models, top-performing text embedding models, and large language models, respectively. Detailed introductions of these models are presented in \S\ref{sec:app_language_models}.

\subsection{Implementation Details}
To construct the architecture presented in Figure~\ref{fig:combined_model}, the function $e^{(t)}$ is implemented as a fully connected layer, while the functions $e^{(n)}_1$, $e^{(n)}_2$, $f_1$, and $f_2$ are implemented using Multi-Layer Perceptron (MLP) structures. The number of layers in each MLP is selected from $\{2, 3, 4\}$, with hidden layer sizes chosen from $\{16, 32, 64\}$. Both attention mechanisms consist of a single layer, with the number of attention heads chosen from $\{1, 3, 6\}$. The hidden dimensions for the query, key, and value vectors in the attention layers are selected from $\{32, 64, 128\}$. A grid search is performed on the validation set to identify the optimal combination of hyper-parameters.

We conducted experiments using both 2080Ti and A100 GPUs. Training a model requires 32 hours on a 2080Ti GPU or 7.5 hours on an A100 GPU. Once trained, the models can handle test cases spanning over a month, requiring only 10.16 seconds for a single prediction. Given the common usage of A100 and superior GPUs in the industry, our computing requirements are practical for real-world applications.

\subsection{Baseline Methods}
\label{sec:baselines}
We evaluate FININ against seven advanced news-based market prediction approaches, representing various techniques for leveraging financial news to predict market movements:

\noindent\textbf{News Appearance Frequency (NAF)}~\cite{huynh2017stock} quantifies news influence based on its frequency of occurrence.

\noindent\textbf{TRNA-based Sentiment Indicator}~\cite{wang2018combining,mohan2019stock,jing2021hybrid}: Many works create market sentiment indicators from news sentiments and use deep learning structures like MLP and LSTM to predict market trends. We experiment with different sentiment indicators and prediction structures based on sentiments from our TRNA dataset to cover these methods. We report the best results as representative of these methods.

\noindent\textbf{FinBERT-based Sentiment Indicator}~\cite{liu2021finbert}: We obtain sentiment scores using FinBERT and use the same methods as the previous baseline to obtain results.

\noindent\textbf{Ensemble Deep Learning Model}~\cite{li2022novel} integrates sentiment analysis with a two-level blending ensemble model, effectively capturing time series events.

\noindent\textbf{FPT}~\cite{zhou2023one} is an advanced pre-training approach for time series analysis, achieving state-of-the-art results across various tasks.

\noindent\textbf{GPT}~\cite{lopez2023can}: A recent study examines the potential of LLMs in predicting market returns using news headlines. They found GPT models obtain the highest Sharpe ratio. 

\noindent\textbf{PEN}~\cite{li2023pen}: PEN uses both textual and price data for market prediction with explainability.

\subsection{Training and Validation}
Our experimental setup considers input data time spans of 1, 3, 5, 10 and 20 trading days, with the 5-day and 20-day settings equivalent to weekly and monthly predictions. Time series cross-validation is used to determine optimal hyper-parameters. The dataset is divided into 10 sliding windows based on dates. Each window contains data from 500 consecutive days. We initiate by collecting the first window from the initial day and then progress by 391 days to obtain the second window. In each window, data is allocated into training, validation, and testing sets in an 8:1:1 ratio, following the chronological order. Reported results are averages across the 10 subsets, providing a robust and reliable evaluation of the model's performance across different time periods and market conditions.

\subsection{Evaluation Metrics}
\label{sec:metrics}
The evaluation of our model is based on three metrics: \textbf{Accuracy (Acc)}, \textbf{Profit and Loss (PnL)} and \textbf{Sharpe ratio (SR)}. These metrics are commonly used in market prediction literature~\cite{ye2020reinforcement, yuemei2021predicting}. PnL measures the cumulative profit or loss generated by the model's predictions, providing a direct financial outcome of its performance. SR measures the risk adjusted returns, quantifying the amount of return achieved per unit of risk. Among them, SR is considered the most important, as it incorporates both returns and risk, which are crucial factors in stock investment. Therefore, when comparing different methods, we prioritize the comparison of SR. PnL serves as a secondary metric. While accuracy is considered, it is given less weight, as high accuracy does not guarantee high profitability. Detailed discussions of PnL and SR are presented in \S\ref{sec:app_metrics}

\section{Results and Discussions}
\label{sec:results_discussions}

\subsection{Language Model Comparison}
To identify the LM that generates the most suitable text embeddings for market prediction, we trained FININ with different LMs as the text encoder for the S\&P 500 market. The results in Table~\ref{tab:lm_results} show RoBERTa outperforming other models across most settings. This suggests that general improvements in model design, as seen in RoBERTa, may be more beneficial than domain-specific adaptations (FinBERT) for news-based market prediction. This observation  aligns with our intuition that financial terms in news are generally understandable, making domain-specific adaptation less crucial.

\begin{table}[b]
\renewcommand\arraystretch{1}
\small
\vspace{-0.4cm}
\begin{center}
\begin{tabular}[b]{c@{\hspace{4pt}} c@{\hspace{4pt}} c@{\hspace{6pt}} c@{\hspace{10pt}} c@{\hspace{10pt}} c@{\hspace{10pt}}}
\toprule
\multirow{2}{*}{\shortstack{Time\\Length}}  & \multirow{2}{*}{Merics}
& \multicolumn{4}{c}{Language Models} \\
\cmidrule(r){3-6}
&\quad
& RoBERTa & FinBERT & BGE & Llama2\\[-0.7mm]
\cmidrule(r){1-2}
\multirow{3}{*}{1-day} 
    & Acc & 0.518 & 0.524 & 0.539 & 0.539  \\
    & PnL & 0.033 & 0.040 & 0.042 & 0.033  \\
    & SR  & \textbf{1.223} & 1.170 & 0.986 & 1.082  \\[-0.7mm]
\cmidrule(r){1-2}
\multirow{3}{*}{3-day} 
    & Acc & 0.537 & 0.524 & 0.518 & 0.535 \\
    & PnL & 0.045 & 0.039 & 0.046 & 0.037 \\
    & SR  & 1.247 & 1.256 & \textbf{1.300} & 1.187 \\[-0.7mm]
\cmidrule(r){1-2}
\multirow{3}{*}{5-day} 
    & Acc & 0.533 & 0.543 & 0.524 & 0.512 \\
    & PnL & 0.032 & 0.019 & 0.019 & 0.037 \\
    & SR  & \textbf{1.192} & 0.948 & 0.462 & 1.131 \\[-0.7mm]
\cmidrule(r){1-2}
\multirow{3}{*}{10-day} 
    & Acc & 0.531 & 0.522 & 0.545 & 0.543 \\
    & PnL & 0.053 & 0.054 & 0.047 & 0.030 \\
    & SR  & \textbf{1.643} & 1.465 & 1.330 & 1.164 \\[-0.7mm]
\cmidrule(r){1-2}
\multirow{3}{*}{20-day} 
    & Acc & 0.522 & 0.520 & 0.520 & 0.551 \\
    & PnL & 0.041 & 0.063 & 0.041 & 0.046 \\
    & SR  & \textbf{1.772} & 1.658 & 1.432 & 1.408 \\
\bottomrule
\end{tabular}
\vspace{-0.5cm}
\captionof{table}{Results of using different LMs in FININ.}
\label{tab:lm_results}
\end{center}
\vspace{-0.3cm}
\end{table}

\begin{table*}[t]
\renewcommand\arraystretch{1}
\vspace{-0.4cm}
\begin{center}
\small
\begin{tabular}[b]{l@{\hspace{4pt}}
c@{\hspace{3.5pt}} c@{\hspace{3.5pt}} c@{\hspace{9pt}} 
c@{\hspace{3.5pt}} c@{\hspace{3.5pt}} c@{\hspace{9pt}}  
c@{\hspace{3.5pt}} c@{\hspace{3.5pt}} c@{\hspace{9pt}}  
c@{\hspace{3.5pt}} c@{\hspace{3.5pt}} c@{\hspace{9pt}} 
c@{\hspace{3.5pt}} c@{\hspace{3.5pt}} c }
\toprule
\multirow{2}{*}{\shortstack{Method}} 
% & \multicolumn{15}{c}{Results} \\
& & 1-day & & & 3-day & & & 5-day & & & 10-day & & & 20-day & \\
& Acc & PnL & SR & Acc & PnL & SR & Acc & PnL & SR & Acc & PnL & SR & Acc & PnL & SR \\
\cmidrule(r){1-16}
\multicolumn{16}{l}{Results of the S\&P 500 Market (Always Buy strategy results in Acc, PnL, SR of 0.520, 0.039, 0.556, respectively.)} \\
\cmidrule(r){1-16}
NAF & 0.522 & 0.017 & 0.678 & 0.520 & 0.018 & 0.744 & 0.520 & -0.015 & -0.076 & 0.500 & 0.016 & 0.480 & 0.522 & 0.024 & 0.644 \\
Sentiments & 0.518 & \textbf{0.036} & 0.914 & \textbf{0.541} & 0.037 & 0.943 & 0.524 & 0.022 & 0.509 & 0.525 & 0.038 & 1.031 & \textbf{0.554} & \textbf{0.045} & 1.085 \\
FinBERT & 0.512 & 0.010 & 0.267 & 0.520 & 0.016 & 0.697 & 0.520 & -0.076 & -0.076 & 0.518 & 0.010 & 0.613 & 0.529 & 0.012 & 0.644 \\
Ensemble & 0.522 & 0.018 & 0.886 & 0.514 & 0.031 & 0.672 & 0.510 & 0.031 & 0.824 & \textbf{0.534} & 0.031 & 1.343 & 0.545 & 0.028 & 0.613 \\
FPT & 0.508 & 0.002 & 0.323 & 0.524 & 0.033 & 0.917 & 0.520 & 0.025 & 1.169 & 0.528 & 0.025 & 0.969 & 0.547 & 0.035 & 0.769 \\
GPT-3.5 & \textbf{0.522} & 0.022 & 0.914 & 0.530 & 0.030 & 1.133 & 0.528 & 0.014 & 0.946 & 0.522 & 0.014 & 0.765 & 0.504 & 0.022 & 1.110 \\
GPT-4 & 0.504 & 0.032 & 0.696 & 0.520 & 0.008 & 0.194 & 0.516 & -0.004 & -0.104 & 0.484 & -0.004 & 0.424 & 0.494 & 0.012 & 0.461 \\
PEN & 0.504 & 0.004 & -0.224 & 0.510 & 0.018 & 0.062 & 0.516 & \textbf{0.034} & 0.457 & 0.531 & 0.048 & 1.167 & 0.539 & 0.015 & 0.625 \\
FININ & 0.518 & 0.033 & \textbf{1.223} & 0.537 & \textbf{0.045} & \textbf{1.247} & \textbf{0.533} & 0.032 & \textbf{1.192} & 0.531 & \underline{\textbf{0.053}} & \textbf{1.643} & 0.522 & 0.014 & \underline{\textbf{1.772}} \\

\cmidrule(r){1-16}
\multicolumn{16}{l}{Results of the NASDAQ 100 Market (Always Buy strategy results in Acc, PnL, SR of 0.518, 0.043, 0.759, respectively.)} \\
\cmidrule(r){1-16}
NAF & 0.537 & 0.014 & 0.589 & 0.527 & 0.002 & 0.299 & 0.555 & 0.021 & 0.816 & 0.541 & 0.021 & 0.853 & 0.522 & 0.016 & 0.648 \\
Sentiments & 0.512 & 0.025 & 0.706 & 0.531 & 0.022 & 0.620 & 0.551 & 0.007 & 0.439 & 0.549 & 0.031 & 0.847 & 0.551 & 0.040 & 1.084 \\
FinBERT	& 0.514 & 0.000 & 0.012 & 0.529 & -0.003 & 0.157 & 0.557 & 0.015 & 0.686 & 0.545 & 0.028 & 0.755 & 0.539 & 0.008 & 0.563 \\
Ensemble & 0.540 & -0.006 & 0.107 & 0.534 & 0.007 & 0.628 & 0.546 & 0.010 & 0.627 & 0.532 & 0.020 & 0.985 & 0.542 & 0.024 & 1.095 \\
FPT & 0.536 & 0.009 & 0.523 & 0.537 & 0.003 & 0.457 & 0.549 & 0.009 & 0.766 & 0.540 & 0.017 & 0.837 & 0.506 & 0.003 & 0.322 \\
GPT-3.5 & 0.538 & 0.010 & \textbf{0.795} & 0.522 & 0.023 & 0.882 & 0.526 & 0.015 & 0.692 & 0.530 & 0.017 & 0.896 & 0.522 & 0.009 & 0.639 \\
PEN & 0.504 & -0.001 & -0.043 & 0.529 & 0.018 & 0.572 & 0.455 & -0.024 & -0.630 & 0.492 & 0.025 & 0.258 & 0.551 & 0.045 & 1.108 \\
FININ & \textbf{0.541} & \textbf{0.025} & 0.651 & \textbf{0.559} & \textbf{0.027} & \textbf{0.966} & \textbf{0.557} & \textbf{0.026} & \textbf{1.095} & \textbf{0.571} & \textbf{0.041} & \textbf{1.307} & \textbf{0.553} & \underline{\textbf{0.046}} & \underline{\textbf{1.449}} \\
\bottomrule
\end{tabular}
\end{center}
\vspace{-0.4cm}
\captionof{table}{Market prediction results for the S\&P 500 and NASDAQ 100 indices. The ``Always Buy'' strategy is a simple benchmark, assuming that prices will always increase the next day, providing basic results of market performance. Bold numbers indicate the highest value for each metric within a specific setting, while underlined numbers denote the highest PnL and SR across all settings. Results for GPT-4 on the NASDAQ 100 are exclusive due to its comparatively lower performance on the S\&P 500 dataset and limited availability compared to GPT-3.5.}
\label{tab:main_results}
\vspace{-0.4cm}
\end{table*}

While BGE exhibits strong performance on many NLP tasks, its results in our financial context are less impressive, reinforcing the notion that no single embedding dominates all tasks~\cite{muennighoff2022mteb}. Llama2, representing advanced LLMs, also shows lower performance. As LLMs are primarily optimized for text generation, our findings reflect the concerns regarding their limitations in text encoding~\cite{jiang2023scaling}. Additionally, Llama2's large embedding size (4096) compared to the typical size (around 1000) introduces challenges for models of general sizes. 

In conclusion, these results highlight the importance of LM selection in effectively encoding market-related textual data. Among the evaluated LMs, RoBERTa exhibited robust overall performance, rendering it the most suitable choice for our FININ model. Subsequent experiments will leverage RoBERTa as the text encoder.

\subsection{Market Prediction Comparison}
\label{sec:prediction_results}

Table~\ref{tab:main_results} presents the market prediction results of the FININ model compared to baseline methods. While the rankings of three metrics do not strongly correlate, the SR is considered the most practical and important one, as mentioned in \S~\ref{sec:metrics}. Despite not always achieving the best Acc or PnL, FININ consistently outperforms all baselines in terms of the SR across different settings.

Furthermore, for both datasets, the best PnL and SR results (underlined) are generated by FININ. FININ's highest SR results, 1.772 on the S\&P 500 and 1.449 on the NASDAQ 100, outperform the best SR achieved by other methods by at least 0.429 and 0.341, respectively, for each dataset.

\noindent\textbf{Discussion on News Use} The comparison highlights FININ's superiority in leveraging extensive news data for market prediction. While baseline methods also incorporate news information, FININ offers two distinct advantages. Firstly, FININ uses both news sentiments and textual headlines. This contrasts with baseline methods that primarily rely on sentiment analysis or text processing. By incorporating multi-modal inputs, FININ can capture interactions between news items, an aspect that other methods overlook. Secondly, FININ's analysis of individual news items enables efficient management of large volumes of news. Baseline methods treat all news items equally, causing influential news to be overwhelmed by general news when faced with substantial data. FININ encodes each news item individually, allowing them to contribute differently to the prediction based on their varying influence. This capability is crucial when dealing with large news volumes, where lack of proper news aggregation in other methods can even impair predictions (compared to the ``Always Buy'' strategy). 
FININ's capability in modeling news interactions is analyzed in detail through case studies in \S\ref{sec:app_case_study}. In summary, FININ's advanced input processing and effective news management techniques position it as a superior tool for using financial news data in market prediction tasks. 

\vspace{-0.12cm}
\noindent\textbf{Discussion on Long Memory Effect and LLM} Our findings reveal the long memory effect of financial news. An increasing trend in the Sharpe ratio is evident as the timeframe of considered news data extends. This trend is particularly pronounced in FININ's results due to its stable and effective news data use. Baseline methods also suggest improvement with extended timeframes, indicating that including data from a broader range of trading days can enhance the predictions.
However, GPT-based methods exhibit an exception. GPT-3.5 and GPT-4 achieve their best results in the 3-day and 1-day prediction settings on the S\&P 500 dataset, with performance decreasing when more days are considered. Examining GPT's outputs revealed its limitations in analyzing news for a complex index like the S\&P 500. In the GPT-based methods, GPT is queried about the positive, negative, or unknown influence of a news item on a company's stock. However, when applied to our datasets, GPT struggles with analyzing the complexities of the S\&P 500 market. 
GPT appears overly cautious in its responses, with a significant portion of news items (55.38\% for GPT-3.5 and 66.78\% for GPT-4) classified as ``unknown''. Consequently, as we consider news from more days, an increasing number of news items contribute nothing to the predictions. Furthermore, the significantly biased news sentiments impair predictions, leading to weak results.

\subsection{Ablation Studies}
\label{sec:ablation_study}

\begin{table}[t]
\renewcommand\arraystretch{1}
\begin{center}
\small
% \vspace{-0.2cm}
\begin{tabular}[b]{c@{\hspace{4pt}} c@{\hspace{4pt}} c@{\hspace{4pt}} c@{\hspace{4pt}} c@{\hspace{4pt}} c}
\toprule
\multirow{4}{*}{\shortstack{Time\\Length}}  & \multirow{4}{*}{Metrics}
& \multicolumn{4}{c}{Model Structure} \\[-0.7mm]
\cmidrule(r){3-6}
& & \scriptsize Complete & \scriptsize FININ-  & \scriptsize FININ-  & \scriptsize FININ-  \\[-0.7mm]
& & \scriptsize FININ  & \scriptsize NTP-NNP & \scriptsize MTP-NTP & \scriptsize MIQ \\[-1mm]
& & \scriptsize  & \tiny (no news data) & \tiny (no textual data) & \tiny (no quantifier) \\[-0.7mm]
\cmidrule(r){1-2}
\multirow{3}{*}{1-day} 
    & Acc & 0.518 & 0.520 & 0.543 & 0.529 \\
    & PnL & 0.033 & 0.034 & 0.033 & 0.015 \\
    & SR  & \textbf{1.223} & 1.021 & 1.093 & 0.741 \\[-0.7mm]
\cmidrule(r){1-2}
\multirow{3}{*}{3-day} 
    & Acc & 0.537 & 0.512 & 0.541 & 0.512 \\
    & PnL & 0.045 & 0.028 & 0.022 & 0.010 \\
    & SR  & \textbf{1.247} & 0.905 & 0.696 & 0.084 \\[-0.7mm]
\cmidrule(r){1-2}
\multirow{3}{*}{5-day} 
    & Acc & 0.533 & 0.506 & 0.504 & 0.529 \\
    & PnL & 0.032 & -0.020 & 0.020 & -0.005 \\
    & SR  & \textbf{1.192} & -0.350 & 0.435 & -0.231 \\[-0.7mm]
\cmidrule(r){1-2}
\multirow{3}{*}{10-day} 
    & Acc & 0.531 & 0.518 & 0.518 & 0.512 \\
    & PnL & 0.053 & 0.025 & 0.039 & 0.021 \\
    & SR  & \textbf{1.643} & 1.120 & 1.301 & 0.967 \\[-0.7mm]
\cmidrule(r){1-2}
\multirow{3}{*}{20-day} 
    & Acc & 0.522 & 0.537 & 0.531 & 0.504 \\
    & PnL & 0.041 & 0.030 & 0.037 & 0.025 \\
    & SR  & \textbf{1.772} & 1.126 & 1.480 & 1.064 \\
\bottomrule
\end{tabular}
\vspace{-0.3cm}
\captionof{table}{Ablation study results of FININ.}
\label{tab:ablation_results}
\end{center}
\vspace{-0.6cm}
\end{table}

To isolate the impact of different data sources and processing methods within FININ, we decompose FININ into five key components: four subsections of the data fusion encoder, the Market Textual data Processor (MTP), the Market Numerical data Processor (MNP), the News Textual data Processor (NTP), and the News Numerical data Processor (NNP), as well as the Market-aware Influence Quantifier (MIQ). We examined three ablated variants: (1) removing news data (FININ-NTP-NNP), (2) removing text inputs (FININ-MTP-NTP), and (3) removing the influence quantifier (FININ-MIQ). When removing MIQ, we aggregated news features via simple averaging instead of using MIQ's scores. 

Table~\ref{tab:ablation_results} presents the ablation study results on the S\&P 500 market. The performance drop in each ablation setup, compared to the complete FININ model, underlines the critical role of news inputs, text data processing, and the news influence quantifier in enhancing market prediction.

The ``FININ-NTP-NNP'' column reveals that predictions based solely on historical price data offer limited long-term forecasting capability, as performance improves marginally with increasing input timeframes. This suggests that the primary information of daily price fluctuations is incorporated into subsequent prices. Incorporating news information, as in the complete FININ model, improves prediction, especially for longer-term forecasts. This finding highlights the heightened relevance of news for capturing long-term stock price dynamics.

The ``FININ-MTP-NTP'' outperforms ``FININ-NTP-NNP'' in most settings, indicating that news sentiment analysis captures additional information not yet reflected in prices. However, the comparison with ``Complete FININ'' shows that relying solely on sentiment analysis misses crucial information within news text, highlighting the necessity of our text processing component.

The ``FININ-MIQ'' results emphasize the significant contribution of the news influence quantifier. Directly averaging news features performs even worse than models without news data (FININ-NTP-NNP). This indicates that simply including a large amount of news data can be detrimental without proper use. The market-aware influence quantifier's scores are essential to extract the predictive power of news for market prediction.

In summary, while market data provides baseline results, incorporating news sentiment and textual information through FININ's processing components yields substantial performance gains. Moreover, FININ's market-aware influence quantifier is crucial for leveraging news data, as it effectively integrates the complex inputs and extracts their predictive power for enhanced market prediction.

\subsection{Insights into Financial News}
Our comprehensive experimental results offer three key insights into the interplay between financial news and market movements, highlighting the need for improved approaches to using news data:

\noindent\textbf{Delayed Market Pricing of News Information}: Despite the S\&P 500 and NASDAQ 100 being highly efficient markets, our study reveals an evident delay in fully integrating news into asset prices. Results and discussions in \S\ref{sec:prediction_results} and \S\ref{sec:ablation_study} prove this through improved performance when using data from previous days. This presents an opportunity for investors to exploit market inefficiencies by incorporating news data into their decision-making.

\noindent\textbf{Long Memory Effect of Financial News}: Our results across various time horizons highlight the sustained impact of news, aligning with the information diffusion hypothesis. While immediate effects of news are apparent in 1-day predictions, improvement is observed across all tested time settings, indicating a delayed influence that persists rather than being released at a specific timeframe.

\noindent\textbf{Limitations of Market Sentiment Analysis}: Relying solely on sentiment analysis, a commonly used approach, may not capture the full complexity of the market's reaction to news. Our findings advocate for a more holistic approach that considers both sentiment and text for a nuanced understanding of news impact on asset prices.

\section{Conclusions}
This paper introduces FININ, a novel model designed to use financial news for market prediction. FININ effectively uses market-related information from news data, outperforming various advanced market prediction methods. Our experiments shed light on the relationships between news and the market, highlighting delayed market pricing of news, the long memory effect of financial news, and the limitations of market sentiment analysis in extracting predictive power from news.

\section*{Limitations}
FININ is a novel approach to building news interactions for market prediction. However, as the related financial theory is less considered in deep learning, there is significant room for further improvement. We discuss the primary limitations of our model as follows:

First, similar to many models using attention mechanisms~\cite{achiam2023gpt, touvron2023llama}, FININ is limited by the attention window size. Attention mechanisms cannot be applied to sequences of unlimited length. Consequently, while it would be ideal to capture detailed interactions among all available news reports, FININ currently focuses on linking news within the same day. The relationship between the information from different days is built by the prediction model. This limitation primarily arises due to the substantial daily news volume (average 702, maximum 1953 per trading day), which poses a challenge for the attention mechanism. Although techniques exist to potentially enlarge the attention window size, this paper prioritizes providing a practical method for building news interactions. Future work can explore methods for capturing long-range news dependencies that extend beyond a single day, which could potentially improve the model's performance.

Second, our proposed FININ model primarily focuses on news data due to the well-established financial theories that demonstrate news' connections to market movements. However, market movements are influenced by a multitude of factors beyond news. Social media information and company reports are just two examples of additional data sources that likely interact with news and price data, but the processing of such datasets is beyond the scope of this paper. FININ has the potential to be extended to incorporate these relevant data sources, potentially leading to a more comprehensive understanding of market information diffusion, price discovery, and predictions.

\section*{Acknowledgments}
This work was supported by the UKRI Centre for Doctoral Training (CDT) in Natural Language Processing through UKRI grant EP/S022481/1. We would like to thank Chang Luo, Guojun Wu, Yftah Ziser, Zheng Zhao, and the anonymous reviewers for their valuable feedback.

\bibliography{ref}

\begin{thebibliography}{53}
\providecommand{\natexlab}[1]{#1}

\bibitem[{Achiam et~al.(2023)Achiam, Adler, Agarwal, Ahmad, Akkaya, Aleman, Almeida, Altenschmidt, Altman, Anadkat et~al.}]{achiam2023gpt}
Josh Achiam, Steven Adler, Sandhini Agarwal, Lama Ahmad, Ilge Akkaya, Florencia~Leoni Aleman, Diogo Almeida, Janko Altenschmidt, Sam Altman, Shyamal Anadkat, et~al. 2023.
\newblock Gpt-4 technical report.
\newblock \emph{arXiv preprint arXiv:2303.08774}.

\bibitem[{Agarwal et~al.(2019)Agarwal, Kumar, and Goel}]{agarwal2019stock}
Shweta Agarwal, Shailendra Kumar, and Utkarsh Goel. 2019.
\newblock Stock market response to information diffusion through internet sources: A literature review.
\newblock \emph{International Journal of Information Management}, 45:118--131.

\bibitem[{Albert~Jr and Smaby(1996)}]{albert1996market}
Robert~L Albert~Jr and Timothy~R Smaby. 1996.
\newblock Market response to analyst recommendations in the “dartboard” column: the information and price-pressure effects.
\newblock \emph{Review of Financial Economics}, 5(1):59--74.

\bibitem[{Araci(2019)}]{araci2019finbert}
Dogu Araci. 2019.
\newblock {FinBERT}: Financial sentiment analysis with pre-trained language models.
\newblock \emph{arXiv preprint arXiv:1908.10063}.

\bibitem[{Barber and Odean(2008)}]{barber2008all}
Brad~M Barber and Terrance Odean. 2008.
\newblock All that glitters: The effect of attention and news on the buying behavior of individual and institutional investors.
\newblock \emph{The review of financial studies}, 21(2):785--818.

\bibitem[{Bekiros et~al.(2017)Bekiros, Nguyen, Junior, and Uddin}]{bekiros2017information}
Stelios Bekiros, Duc~Khuong Nguyen, Leonidas~Sandoval Junior, and Gazi~Salah Uddin. 2017.
\newblock Information diffusion, cluster formation and entropy-based network dynamics in equity and commodity markets.
\newblock \emph{European Journal of Operational Research}, 256(3):945--961.

\bibitem[{Calomiris and Mamaysky(2019)}]{calomiris2019news}
Charles~W Calomiris and Harry Mamaysky. 2019.
\newblock How news and its context drive risk and returns around the world.
\newblock \emph{Journal of Financial Economics}, 133(2):299--336.

\bibitem[{Cheung et~al.(2019)Cheung, Fatum, and Yamamoto}]{cheung2019exchange}
Yin-Wong Cheung, Rasmus Fatum, and Yohei Yamamoto. 2019.
\newblock The exchange rate effects of macro news after the global financial crisis.
\newblock \emph{Journal of International Money and Finance}, 95:424--443.

\bibitem[{Christensen and Nielsen(2007)}]{christensen2007effect}
Bent~Jesper Christensen and Morten~{\O}rregaard Nielsen. 2007.
\newblock The effect of long memory in volatility on stock market fluctuations.
\newblock \emph{The Review of Economics and Statistics}, 89(4):684--700.

\bibitem[{Dewally(2003)}]{dewally2003internet}
Michael Dewally. 2003.
\newblock Internet investment advice: Investing with a rock of salt.
\newblock \emph{Financial Analysts Journal}, 59(4):65--77.

\bibitem[{Ding et~al.(2020)Ding, Wu, Sun, Guo, and Guo}]{ding2020hierarchical}
Qianggang Ding, Sifan Wu, Hao Sun, Jiadong Guo, and Jian Guo. 2020.
\newblock Hierarchical multi-scale gaussian transformer for stock movement prediction.
\newblock In \emph{IJCAI}, pages 4640--4646.

\bibitem[{Dong et~al.(2020)Dong, Yan, Almudaifer, Yan, Jiang, and Zhou}]{dong2020belt}
Yingzhe Dong, Da~Yan, Abdullateef~Ibrahim Almudaifer, Sibo Yan, Zhe Jiang, and Yang Zhou. 2020.
\newblock Belt: A pipeline for stock price prediction using news.
\newblock In \emph{2020 IEEE International Conference on Big Data (Big Data)}, pages 1137--1146. IEEE.

\bibitem[{Fama(1970)}]{fama1970efficient}
Eugene~F Fama. 1970.
\newblock Efficient capital markets: A review of theory and empirical work.
\newblock \emph{The journal of Finance}, 25(2):383--417.

\bibitem[{Fischer and Krauss(2018)}]{fischer2018deep}
Thomas Fischer and Christopher Krauss. 2018.
\newblock Deep learning with long short-term memory networks for financial market predictions.
\newblock \emph{European journal of operational research}, 270(2):654--669.

\bibitem[{Hirshleifer and Sheng(2022)}]{hirshleifer2022macro}
David Hirshleifer and Jinfei Sheng. 2022.
\newblock Macro news and micro news: complements or substitutes?
\newblock \emph{Journal of Financial Economics}, 145(3):1006--1024.

\bibitem[{Ho et~al.(2013)Ho, Shi, and Zhang}]{ho2013does}
Kin-Yip Ho, Yanlin Shi, and Zhaoyong Zhang. 2013.
\newblock How does news sentiment impact asset volatility? evidence from long memory and regime-switching approaches.
\newblock \emph{The North American Journal of Economics and Finance}, 26:436--456.

\bibitem[{Hou et~al.(2021)Hou, Xu, Liu, Liu, Bian, Wu, Li, Chen, and Liu}]{hou2021stock}
Min Hou, Chang Xu, Yang Liu, Weiqing Liu, Jiang Bian, Le~Wu, Zhi Li, Enhong Chen, and Tie-Yan Liu. 2021.
\newblock Stock trend prediction with multi-granularity data: A contrastive learning approach with adaptive fusion.
\newblock In \emph{Proceedings of the 30th ACM International Conference on Information \& Knowledge Management}, pages 700--709.

\bibitem[{Hu et~al.(2019)Hu, Wang, Liu, Ji, Chen, Zhao, Ma, and Yan}]{hu2019transformation}
Wenpeng Hu, Mengyu Wang, Bing Liu, Feng Ji, Haiqing Chen, Dongyan Zhao, Jinwen Ma, and Rui Yan. 2019.
\newblock Transformation of dense and sparse text representations.
\newblock \emph{arXiv preprint arXiv:1911.02914}.

\bibitem[{Hu et~al.(2020)Hu, Wang, Qin, Ma, and Liu}]{hu2020hrn}
Wenpeng Hu, Mengyu Wang, Qi~Qin, Jinwen Ma, and Bing Liu. 2020.
\newblock Hrn: A holistic approach to one class learning.
\newblock \emph{Advances in neural information processing systems}, 33:19111--19124.

\bibitem[{Huynh and Smith(2017)}]{huynh2017stock}
Thanh~D Huynh and Daniel~R Smith. 2017.
\newblock Stock price reaction to news: The joint effect of tone and attention on momentum.
\newblock \emph{Journal of Behavioral Finance}, 18(3):304--328.

\bibitem[{Jiang et~al.(2023)Jiang, Huang, Luan, Wang, and Zhuang}]{jiang2023scaling}
Ting Jiang, Shaohan Huang, Zhongzhi Luan, Deqing Wang, and Fuzhen Zhuang. 2023.
\newblock Scaling sentence embeddings with large language models.
\newblock \emph{arXiv preprint arXiv:2307.16645}.

\bibitem[{Jiang(2021)}]{jiang2021applications}
Weiwei Jiang. 2021.
\newblock Applications of deep learning in stock market prediction: recent progress.
\newblock \emph{Expert Systems with Applications}, 184:115537.

\bibitem[{Jing et~al.(2021)Jing, Wu, and Wang}]{jing2021hybrid}
Nan Jing, Zhao Wu, and Hefei Wang. 2021.
\newblock A hybrid model integrating deep learning with investor sentiment analysis for stock price prediction.
\newblock \emph{Expert Systems with Applications}, 178:115019.

\bibitem[{Kerl and Walter(2007)}]{kerl2007market}
Alexander~G Kerl and Andreas Walter. 2007.
\newblock Market responses to buy recommendations issued by personal finance magazines: effects of information, price-pressure, and company characteristics.
\newblock \emph{Review of Finance}, 11(1):117--141.

\bibitem[{Krauss et~al.(2017)Krauss, Do, and Huck}]{krauss2017deep}
Christopher Krauss, Xuan~Anh Do, and Nicolas Huck. 2017.
\newblock Deep neural networks, gradient-boosted trees, random forests: Statistical arbitrage on the s\&p 500.
\newblock \emph{European Journal of Operational Research}, 259(2):689--702.

\bibitem[{Li(2010)}]{li2010information}
Feng Li. 2010.
\newblock The information content of forward-looking statements in corporate filings—a na{\"\i}ve bayesian machine learning approach.
\newblock \emph{Journal of Accounting Research}, 48(5):1049--1102.

\bibitem[{Li et~al.(2020)Li, Tan, Wang, and Chen}]{li2020multimodal}
Qing Li, Jinghua Tan, Jun Wang, and Hsinchun Chen. 2020.
\newblock A multimodal event-driven lstm model for stock prediction using online news.
\newblock \emph{IEEE Transactions on Knowledge and Data Engineering}, 33(10):3323--3337.

\bibitem[{Li et~al.(2023)Li, Liao, Chen, and Yan}]{li2023pen}
Shuqi Li, Weiheng Liao, Yuhan Chen, and Rui Yan. 2023.
\newblock Pen: prediction-explanation network to forecast stock price movement with better explainability.
\newblock In \emph{Proceedings of the Thirty-Seventh AAAI Conference on Artificial Intelligence and Thirty-Fifth Conference on Innovative Applications of Artificial Intelligence and Thirteenth Symposium on Educational Advances in Artificial Intelligence}, pages 5187--5194.

\bibitem[{Li and Pan(2022)}]{li2022novel}
Yang Li and Yi~Pan. 2022.
\newblock A novel ensemble deep learning model for stock prediction based on stock prices and news.
\newblock \emph{International Journal of Data Science and Analytics}, pages 1--11.

\bibitem[{Lillo et~al.(2015)Lillo, Miccich{\`e}, Tumminello, Piilo, and Mantegna}]{lillo2015news}
Fabrizio Lillo, Salvatore Miccich{\`e}, Michele Tumminello, Jyrki Piilo, and Rosario~N Mantegna. 2015.
\newblock How news affects the trading behaviour of different categories of investors in a financial market.
\newblock \emph{Quantitative Finance}, 15(2).

\bibitem[{Liu et~al.(2019)Liu, Ott, Goyal, Du, Joshi, Chen, Levy, Lewis, Zettlemoyer, and Stoyanov}]{liu2019roberta}
Yinhan Liu, Myle Ott, Naman Goyal, Jingfei Du, Mandar Joshi, Danqi Chen, Omer Levy, Mike Lewis, Luke Zettlemoyer, and Veselin Stoyanov. 2019.
\newblock {RoBERTa}: A robustly optimized {BERT} pretraining approach.
\newblock \emph{arXiv preprint arXiv:1907.11692}.

\bibitem[{Liu et~al.(2021)Liu, Huang, Huang, Li, and Zhao}]{liu2021finbert}
Zhuang Liu, Degen Huang, Kaiyu Huang, Zhuang Li, and Jun Zhao. 2021.
\newblock Finbert: A pre-trained financial language representation model for financial text mining.
\newblock In \emph{Proceedings of the twenty-ninth international conference on international joint conferences on artificial intelligence}, pages 4513--4519.

\bibitem[{Lopez-Lira and Tang(2023)}]{lopez2023can}
Alejandro Lopez-Lira and Yuehua Tang. 2023.
\newblock Can chatgpt forecast stock price movements? return predictability and large language models.
\newblock \emph{Return Predictability and Large Language Models (April 6, 2023)}.

\bibitem[{Luo et~al.(2023)Luo, Liao, Li, Cheng, and Yan}]{luo2023causality}
Di~Luo, Weiheng Liao, Shuqi Li, Xin Cheng, and Rui Yan. 2023.
\newblock Causality-guided multi-memory interaction network for multivariate stock price movement prediction.
\newblock In \emph{Proceedings of the 61st Annual Meeting of the Association for Computational Linguistics (Volume 1: Long Papers)}, pages 12164--12176.

\bibitem[{Mitra and Mitra(2011)}]{mitra2011applications}
Leela Mitra and Gautam Mitra. 2011.
\newblock Applications of news analytics in finance: A review.
\newblock \emph{The handbook of news analytics in finance}, 596(1).

\bibitem[{Mohan et~al.(2019)Mohan, Mullapudi, Sammeta, Vijayvergia, and Anastasiu}]{mohan2019stock}
Saloni Mohan, Sahitya Mullapudi, Sudheer Sammeta, Parag Vijayvergia, and David~C Anastasiu. 2019.
\newblock Stock price prediction using news sentiment analysis.
\newblock In \emph{2019 IEEE Fifth International Conference on Big Data Computing Service and Applications (BigDataService)}, pages 205--208. IEEE.

\bibitem[{Muennighoff et~al.(2022)Muennighoff, Tazi, Magne, and Reimers}]{muennighoff2022mteb}
Niklas Muennighoff, Nouamane Tazi, Lo{\"\i}c Magne, and Nils Reimers. 2022.
\newblock Mteb: Massive text embedding benchmark.
\newblock \emph{arXiv preprint arXiv:2210.07316}.

\bibitem[{Orimoloye et~al.(2020)Orimoloye, Sung, Ma, and Johnson}]{orimoloye2020comparing}
Larry~Olanrewaju Orimoloye, Ming-Chien Sung, Tiejun Ma, and Johnnie~EV Johnson. 2020.
\newblock Comparing the effectiveness of deep feedforward neural networks and shallow architectures for predicting stock price indices.
\newblock \emph{Expert Systems with Applications}, 139:112828.

\bibitem[{Ray and Tsay(2000)}]{ray2000long}
Bonnie~K Ray and Ruey~S Tsay. 2000.
\newblock Long-range dependence in daily stock volatilities.
\newblock \emph{Journal of Business \& Economic Statistics}, 18(2):254--262.

\bibitem[{Sattler(2013)}]{sattler2013markets}
Thomas Sattler. 2013.
\newblock Do markets punish left governments?
\newblock \emph{The Journal of Politics}, 75(2):343--356.

\bibitem[{Tetlock(2007)}]{tetlock2007giving}
Paul~C Tetlock. 2007.
\newblock Giving content to investor sentiment: The role of media in the stock market.
\newblock \emph{The Journal of finance}, 62(3):1139--1168.

\bibitem[{Touvron et~al.(2023)Touvron, Martin, Stone, Albert, Almahairi, Babaei, Bashlykov, Batra, Bhargava, Bhosale et~al.}]{touvron2023llama}
Hugo Touvron, Louis Martin, Kevin Stone, Peter Albert, Amjad Almahairi, Yasmine Babaei, Nikolay Bashlykov, Soumya Batra, Prajjwal Bhargava, Shruti Bhosale, et~al. 2023.
\newblock Llama 2: Open foundation and fine-tuned chat models.
\newblock \emph{arXiv preprint arXiv:2307.09288}.

\bibitem[{Vaswani et~al.(2017)Vaswani, Shazeer, Parmar, Uszkoreit, Jones, Gomez, Kaiser, and Polosukhin}]{vaswani2017attention}
Ashish Vaswani, Noam Shazeer, Niki Parmar, Jakob Uszkoreit, Llion Jones, Aidan~N Gomez, {\L}ukasz Kaiser, and Illia Polosukhin. 2017.
\newblock Attention is all you need.
\newblock \emph{Advances in neural information processing systems}, 30.

\bibitem[{Wang and Ma(2024)}]{wang2024mana}
Mengyu Wang and Tiejun Ma. 2024.
\newblock Mana-net: Mitigating aggregated sentiment homogenization with news weighting for enhanced market prediction.
\newblock \emph{arXiv preprint arXiv:2409.05698}.

\bibitem[{Wang et~al.(2018)Wang, Xu, and Zheng}]{wang2018combining}
Qili Wang, Wei Xu, and Han Zheng. 2018.
\newblock Combining the wisdom of crowds and technical analysis for financial market prediction using deep random subspace ensembles.
\newblock \emph{Neurocomputing}, 299:51--61.

\bibitem[{Xiao et~al.(2023)Xiao, Liu, Zhang, and Muennighof}]{xiao2023c}
Shitao Xiao, Zheng Liu, Peitian Zhang, and Niklas Muennighof. 2023.
\newblock C-pack: Packaged resources to advance general chinese embedding.
\newblock \emph{arXiv preprint arXiv:2309.07597}.

\bibitem[{Xu et~al.(2021)Xu, Liu, Xu, Bian, Yin, and Liu}]{xu2021rest}
Wentao Xu, Weiqing Liu, Chang Xu, Jiang Bian, Jian Yin, and Tie-Yan Liu. 2021.
\newblock Rest: Relational event-driven stock trend forecasting.
\newblock In \emph{Proceedings of the web conference 2021}, pages 1--10.

\bibitem[{Xu and Cohen(2018)}]{xu2018stock}
Yumo Xu and Shay~B Cohen. 2018.
\newblock Stock movement prediction from tweets and historical prices.
\newblock In \emph{Proceedings of the 56th Annual Meeting of the Association for Computational Linguistics (Volume 1: Long Papers)}, pages 1970--1979.

\bibitem[{Ye et~al.(2020)Ye, Pei, Wang, Chen, Zhu, Xiao, and Li}]{ye2020reinforcement}
Yunan Ye, Hengzhi Pei, Boxin Wang, Pin-Yu Chen, Yada Zhu, Jun Xiao, and Bo~Li. 2020.
\newblock Reinforcement-learning based portfolio management with augmented asset movement prediction states.
\newblock In \emph{AAAI Conference on Artificial Intelligence}. AAAI press.

\bibitem[{Yu et~al.(2019)Yu, Zhao, Tang, and Yang}]{yu2019online}
Lean Yu, Yaqing Zhao, Ling Tang, and Zebin Yang. 2019.
\newblock Online big data-driven oil consumption forecasting with google trends.
\newblock \emph{International Journal of Forecasting}, 35(1):213--223.

\bibitem[{Yuemei et~al.(2021)Yuemei, Zihou, and Zixin}]{yuemei2021predicting}
Xu~Yuemei, Wang Zihou, and Wu~Zixin. 2021.
\newblock Predicting stock trends with cnn-bilstm based multi-feature integration model.
\newblock \emph{Data Analysis and Knowledge Discovery}, 5(7):126--138.

\bibitem[{Zhang et~al.(2016)Zhang, Song, Shen, and Zhang}]{zhang2016market}
Yongjie Zhang, Weixin Song, Dehua Shen, and Wei Zhang. 2016.
\newblock Market reaction to internet news: Information diffusion and price pressure.
\newblock \emph{Economic Modelling}, 56:43--49.

\bibitem[{Zhou et~al.(2023)Zhou, Niu, Wang, Sun, and Jin}]{zhou2023one}
Tian Zhou, Peisong Niu, Xue Wang, Liang Sun, and Rong Jin. 2023.
\newblock One fits all: Power general time series analysis by pretrained lm.
\newblock \emph{arXiv preprint arXiv:2302.11939}.

\end{thebibliography}
\appendix

\section{Data}
\label{sec:app_data}
\noindent \textbf{Market Dataset:}  The S\&P 500 Index, a benchmark for the U.S. market, tracks 500 leading publicly traded companies~\cite{krauss2017deep, fischer2018deep}. The NASDAQ 100 Index focuses on 100 of the largest, most actively traded non-financial companies listed on the Nasdaq Stock Market~\cite{hou2021stock, ding2020hierarchical}. Both indices are renowned for representing the largest and most influential U.S. market segments and are widely used as benchmark datasets for market prediction. We collected daily S\&P 500 and NASDAQ 100 prices from 2003 to 2018, including six key features: open price, close price, adjusted close price, high price, low price, and trading volume, as market numerical data $\boldsymbol{m}^{(n)}_{d}$. Also, we collected the descriptions of these two markets and their constituent companies' names as textual data $\boldsymbol{m}^{(t)}_{d}$.

\noindent \textbf{News Dataset:} Our used TRNA dataset spans from 2003 to 2018. It tracks over 25,000 equities and nearly 40 commodities and energy topics, providing a comprehensive news source for studying financial markets~\cite{mitra2011applications}. The news dataset provides detailed information about collected news, including sentiment scores from an industry-established news-analytics system. These scores indicate the likelihood that the news will have a positive, neutral, or negative influence on the mentioned financial instruments. In our approach, we consider headlines as news textual data $\boldsymbol{r}^{(t)}_{d, i}$ and the sentiment scores as news numerical data $\boldsymbol{r}^{(n)}_{d, i}$. Our experiments aim to encompass all available news items, despite the considerable daily news volume, which exhibits substantial volatility. The maximum daily news volume reaches 1,953, which is roughly 80 times greater than the minimum count of 24. This highlights the dynamic nature of the news volume on a daily basis.

\section{Detailed Experiment Settings}
\label{sec:app_settings}

\begin{table*}[t]
\renewcommand\arraystretch{1}
\small
\vspace{-0.2cm}
\begin{center}
\begin{tabular}[b]{c@{\hspace{4pt}} c@{\hspace{4pt}} l@{\hspace{4pt}} c}
\toprule
Date & Case ID & News Headline & Weight \\
2008-09-26 & 1 & \underline{JPMorgan Chase \& Co <JPM.N> says WaMu clientele very similar to JPMorgan's} & 0.82 \\
2008-09-26 & 2 & UPDATE 4-WaMu is largest US bank failure; JPMorgan buys assets  & 0.98 \\
\cmidrule{1-4}
2008-10-14 & 3 & BANK OF AMERICA <BAC.N> up more than 15 pct as U.S. to inject \$250 bln into banks & 0.99 \\
2008-10-14 & 4 & \underline{STOCKS NEWS US-Banks reacting differently to government plan} & 0.77 \\
\cmidrule{1-4}
2008-10-24 & 5 & \underline{GM to delay unveiling of Buick LaCrosse model} & 0.76 \\
2008-10-24 & 6 & \underline{GM <GM.N> to delay unveiling of Buick LaCrosse, Cadillac CTS coupe models} & 0.71 \\
2008-10-24 & 7 & U.S. bank shares tumble on recession worry & 0.92 \\
2008-10-24 & 8 & Stocks News - Auto companies down, GM intensifies merger talks & 0.83 \\
\bottomrule
\end{tabular}
\vspace{-0.2cm}
\captionof{table}{News Cases with their evaluated weights from FININ.}
\label{tab:case_study}
\end{center}
\vspace{-0.6cm}
\end{table*}

\subsection{Language Models}
\label{sec:app_language_models}
Our candidate LMs encompass a diverse range of architectures, including pre-trained encoder models, financial domain-specific models, top-performing text embedding models, and recent large language models.

\noindent\textbf{RoBERTa}~\cite{liu2019roberta} is a well-known extension of BERT model. It has achieved success in various NLP tasks. Due to its robust performance and versatility, RoBERTa is an indispensable option when doing NLP-related tasks.

\noindent\textbf{FinBERT}~\cite{liu2021finbert} is a domain-specific adaptation of the BERT model. It is designed specifically for financial sentiment analysis. This model benefits from pre-training via multi-task learning on extensive financial corpora, making it particularly effective for analyzing financial texts.

\noindent\textbf{BGE}~\cite{xiao2023c} (BAAI General Embedding) is a leading text encoder model that has achieved top performance on the MTEB (Massive Text Embedding Benchmark) leaderboard~\cite{muennighoff2022mteb}. The MTEB is an extensive benchmark designed to evaluate text embeddings across eight NLP tasks, including 58 datasets and supporting 112 languages. This extensive scope ensures that models like BGE are tested against a wide range of NLP challenges, highlighting the quality of their generated text embeddings.

\noindent\textbf{Llama2}~\cite{touvron2023llama} is an advanced large language model that has the potential to be used for a wide variety of purposes. It has already shown competitive performance on many NLP tasks.

\subsection{Evaluation Metrics}
\label{sec:app_metrics}
\noindent \textbf{PnL} quantifies the cumulative profit or loss experienced by a portfolio during a designated time frame. We calculate the PnL for each trading day and sum up the daily PnL values across the entire testing period. The PnL for all forecasts spanning $D$ days can be expressed as (note it requires trading information from day $D+1$):

\begin{equation}
\label{eq:pnl}
\textit{PnL} = \sum^D_{d=1} \textit{flag}_d \cdot \frac{\boldsymbol{p}^c_{d+1} - \boldsymbol{p}^c_{d}}{\boldsymbol{p}^c_{d}},
\end{equation}
where $\textit{flag}_d=1$ if $\hat{y}_d = y_d$ and $\textit{flag}_d=-1$ otherwise.

\noindent \textbf{Sharpe Ratio} measures the investment performance in relation to a risk-free asset. To compute the daily Sharpe Ratio within the test sets, we use the following formula:
\begin{equation}
\label{eq:sharpe_ratio}
\textit{SR} = \left(\displaystyle\frac{1}{n}\displaystyle\sum_{d=1}^D \displaystyle flag_d\frac{\boldsymbol{p}^c_{d+1} - \boldsymbol{p}^c_{d}}{\boldsymbol{p}^c_{d}} - R_f\right)  \Bigg/ {\sigma(R)},
\end{equation}
where $R_f$ represents the return rate of an investment with zero risks, meaning that it is the return that investors could expect for taking no risk, such as a Treasury bond investment. In this context, we use a value of 0.02 for $R_f$, which corresponds to the average US Treasury rate during our data collection period. Furthermore, $\sigma(R)$ denotes the standard deviation of the excess return of the asset, denoted as $R$.

\section{Case Study}
\label{sec:app_case_study}

To evaluate FININ's ability to capture news interactions, we analyze several news cases from the 2008 financial crisis. This period is marked by significant market events that clearly demonstrate the impact of news interactions on financial markets. We specifically select news events with obvious market repercussions to evaluate FININ's responses, since it is hard for humans to clearly ascertain the market impact of most news pieces, due to the large volume of news and the complexity of their interactions.

Since the news weights are normalized using the softmax function to ensure their sum is one, the range of news weights varies with the daily news volume. In our case study, to make weights from different days comparable, we use min-max normalization to rescale all weights to a consistent range of 0 to 1. Table~\ref{tab:case_study} lists these news reports and their normalized weights.

Given that the daily average normalized weights are around 0.48, these listed reports with weights exceeding 0.7 are highlighted by FININ as having a more significant contribution to the market prediction outcomes. However, when read individually, the underlined news reports seem to describe ordinary company events, similar to numerous general financial news items reported daily. Nonetheless, when considering the broader context of other news events, the importance of these underlined reports becomes more apparent.

On 2008-09-26, News 2 announced the failure of Washington Mutual (WaMu), the largest bank failure at that time, and JPMorgan's acquisition of its assets. This was a highly significant event during the 2008 financial crisis. Consequently, News 1, JPMorgan's commentary on WaMu, carried substantial influence on that day. Similarly, on 2008-10-14, following the announcement of the substantial bank rescue package (News 3), News 4, regarding U.S. banks' reactions to the government's bailout plan, gained prominence. Lastly, on 2008-10-24, discussions surrounding a potential merger involving General Motors (GM) (News 8) and concerns about the U.S. recession (News 7) heightened the impact of news related to GM, including News 5 and News 6, the announcements about product delays. Therefore, while the underlined news reports may seem ordinary when viewed in isolation, their significance becomes evident when considered within the broader market context and major events of the time.

In summary, these case studies demonstrate the effectiveness of FININ's modeling of news interactions. While some news items may seem individually general, FININ recognize their importance and assigns higher weights to them, due to considering their interactions with other breaking news events.

\end{document}